\documentclass[aps,pra,twocolumn,amsmath,amssymb,superscriptaddress]{revtex4-2}

\usepackage{bm}
\usepackage{physics}
\usepackage[utf8]{inputenc}
\usepackage{amsmath,amsthm}
\usepackage{graphicx}
\usepackage[colorlinks = true,
            linkcolor = blue,
            urlcolor  = blue,
            citecolor = blue,
            anchorcolor = blue]{hyperref}
\usepackage{braket}
\usepackage{xcolor}
\begin{document}
\title{Dissipative Dynamics and Active Stabilization of Linear and Nonlinear Waves in Non-PT-Symmetric Harmonic Traps}
\author{Mario Salerno} \affiliation{Dipartimento di Fisica
  ``E.R. Caianiello'', and INFN Sezione di Napoli - Gruppo Collegato di Salerno,
  Universit\`a di Salerno, Via Giovanni Paolo II, 84084 Fisciano (SA),
  Italy}
\date{\today}

\begin{abstract}
We investigate the dissipative dynamics of linear and nonlinear waves in harmonic traps by means of engineered complex non-Hermitian potentials. By combining an analytical mapping between real and complex Schr\"odinger equations with direct numerical simulations, we show that while in the linear case the damped motion leads to the formation of a stationary state at the trap center, in the nonlinear case a static potential design alone is insufficient to ensure long-term stability. Instead, the system relaxes toward a long-lived metastable configuration that eventually undergoes decay or collapse. To overcome this limitation, we introduce a time-dependent modulation of the nonlinearity that effectively converts these metastable states into robust non-equilibrium stationary states. This approach establishes a general strategy for controlling nonlinear waves in non-Hermitian systems, with potential applications in photonics and Bose--Einstein condensates.
\end{abstract}

\maketitle

\section{Introduction}

The study of dissipative wave phenomena in nonlinear media with complex parameters has become a central topic in modern physics~\cite{Aranson2002,Akhmediev2008}. Such systems arise naturally in a wide range of contexts, from the optics of active media, such as laser cavities and amplifying fibers~\cite{Kivshar2003,Lugiato2015}, to quantum systems with effective non-Hermitian dynamics describing interactions with an environment~\cite{Breuer2002,Rotter2009}. In recent years, particular attention has been devoted to nonlinear wave propagation in complex potentials, motivated by advances in matter waves in dissipative optical lattices~\cite{Grynberg2001,malomed2006soliton,brazhnyi2004,Morsch2006} and by the rapid development of non-Hermitian photonics~\cite{ElGanainy2018,Feng2017}.

A major breakthrough in this field was the introduction of ${\cal PT}$-symmetric potentials, which are invariant under the combined action of parity (${\cal P}$) and time-reversal (${\cal T}$) symmetry~\cite{Bender1998,Bender2007}. These systems, extensively investigated theoretically~\cite{Konotop2016,Longhi2009} and experimentally realized in optical settings \cite{Guo2009,Ruter2010}, can exhibit entirely real spectra despite the presence of gain and loss. While the linear properties of ${\cal PT}$-symmetric systems, including phase transitions and exceptional points~\cite{Miri2019}, are now well understood, the nonlinear regime remains comparatively less explored. In particular, although stationary localized and periodic states exist in ${\cal PT}$-symmetric media~\cite{Musslimani2008,Abdullaev2010,Zezyulin2012,Suchkov2016}, their stability and dynamical robustness remain open issues.

This naturally raises a broader question: can stable nonlinear states with real energies exist even in the absence of symmetry? While symmetry plays a crucial role in ensuring real spectra in linear systems, nonlinear media may sustain localized states through a balance between dispersion, nonlinearity, and gain--loss mechanisms, even when the underlying potential is not ${\cal PT}$ symmetric~\cite{Zezyulin2012,Graefe2012}.

A paradigmatic setting to address this problem is the dynamics of a localized wave in a harmonic trap. While linear wave packets can be damped to rest at the trap center, the required complex potential is not a simple imaginary gradient $i\gamma x$. Indeed, a ${\cal PT}$-symmetric potential of this form cannot damp the center-of-mass motion. Efficient damping requires, as we demonstrate below, a non-${\cal PT}$-symmetric complex potential whose imaginary part combines an odd contribution with a quadratic term that explicitly breaks ${\cal PT}$ symmetry.

In this work, we address this problem by combining numerical simulations with an analytical construction based on a mapping between real and complex nonlinear Schr\"odinger equations~\cite{salerno-2013}. This approach allows us to design static non-${\cal PT}$-symmetric potentials that, in the linear case, reproduce the desired center-of-mass damping and lead to the formation of a non-equilibrium stationary state (NESS)~\cite{Prosen2011}.

In the nonlinear regime, however, the situation is fundamentally different. An initial condition displaced from equilibrium undergoes damped motion toward the trap center but does not relax to a truly stationary state. Instead, the dynamics leads to a long-lived metastable configuration whose eventual fate depends on the sign of the nonlinearity: decay in the repulsive case and collapse in the attractive one. This behavior reflects an imperfect balance between gain and loss at the trap center, which cannot be fully compensated by a static potential design alone.

To stabilize the relaxed state at the center of the trap, we introduce a time-dependent nonlinear management  protocol~\cite{abdullaev2003nonlinearity} that dynamically modulates the interaction strength as the wave approaches the metastable regime. This active control suppresses the residual gain--loss imbalance and transforms the metastable configuration into a robust NESS.

To the best of our knowledge, this work provides the first evidence that controlled damping of nonlinear waves toward a stationary state can be achieved by combining static non-${\cal PT}$ complex potentials with active nonlinear control. These results are relevant for non-Hermitian photonics, where dissipative solitons must be stabilized in the presence of gain and loss~\cite{Akhmediev2008}, and for Bose--Einstein condensates, where Feshbach resonance techniques allow precise tuning of interatomic interactions~\cite{pitaevskii2016}.

The paper is organized as follows. In Sec.~II, we introduce the model equations and the mapping between real and complex NLS equations. In Sec.~III, we apply the method to the linear harmonic oscillator, where the mapping yields exact dissipative dynamics. In Sec.~IV, we extend the analysis to the nonlinear regime, considering both repulsive and attractive interactions. In Sec.~V, we demonstrate the stabilization of metastable states via nonlinear management. Finally, Sec.~VI summarizes our conclusions and discusses future perspectives.


\section{Model equations and mapping between real and complex NLS}

The dynamics of nonlinear waves in non-Hermitian media is described by the normalized nonlinear Schr\"odinger (NLS) equation:
\begin{equation}\label{sys}
i\psi_t = -\frac{1}{2}\psi_{xx} + \mathcal{V}(x)\psi + \sigma |\psi|^2 \psi,
\end{equation}
where the complex potential $\mathcal{V}(x)$ accounts for linear gain and loss:
\begin{equation}
\mathcal{V}(x) = V(x) + i W(x),
\end{equation}
with $V(x)$ and $W(x)$ being real functions.

This model encompasses several physically relevant situations. In the linear limit, it describes quantum dissipative systems or light propagation in lossy waveguides. In the nonlinear regime, it corresponds to the Gross--Pitaevskii equation for Bose--Einstein condensates with particle exchange, or to light propagation in active photonic media.

We seek stationary solutions of the form:
\begin{equation}
\psi(x,t) = A(x) e^{i \theta(x)} e^{-i \omega t},
\label{stationary}
\end{equation}
where $A(x)$ and $\theta(x)$ are real functions. Substituting Eq.~(\ref{stationary}) into Eq.~(\ref{sys}) yields:
\begin{eqnarray}
\label{sys2}
\omega A + \frac{1}{2}A_{xx} - \sigma A^3 - \frac{1}{2}(\theta_x)^2 A - V A &=& 0, \\
\label{sys3}
\frac{1}{2} A\theta_{xx} + A_x\theta_x - W A &=& 0.
\end{eqnarray}

Equation~(\ref{sys3}) has the form of a continuity equation for the particle (or optical) current. Multiplying it by $A$, it can be written as:
\begin{equation}
\frac{1}{2} (\theta_x A^2)_x = W A^2 \equiv \frac{dF(x)}{dx},
\label{continuity}
\end{equation}
where $F(x)$ represents the cumulative gain--loss balance. Integration of Eq.~(\ref{continuity}) gives:
\begin{eqnarray} 
\label{tetagen}
\theta_x &=& 2 \frac{F(x)}{A^2}, \\ 
W(x) &=& \frac{1}{A^2} \frac{d F}{dx}.
\label{cmpxpot}
\end{eqnarray}
Substituting Eq.~(\ref{tetagen}) into Eq.~(\ref{sys2}) yields a nonlinear eigenvalue problem for the amplitude:
\begin{equation}
\label{eigenval-prob-gen}
\left\{-\frac{1}{2} \frac{d^2}{dx^2} + V(x) + \sigma A^2 + 2 \left(\frac{F(x)}{A^2} \right)^2 \right\} A = \omega A,
\end{equation}
which is equivalent to the original complex NLS for stationary states.

The mapping becomes particularly useful when $F(x)$ is chosen as a functional of the density. In the simplest case, we take:
\begin{equation}
F(x) = \frac{1}{2} C_n A^{n+2}, \quad n=0,1,2,\dots,
\label{complex}
\end{equation}
where $C_n$ is a real constant. Equation~(\ref{eigenval-prob-gen}) then reduces to the real eigenvalue problem:
\begin{equation}
\label{eigenval-prob}
\left\{-\frac{1}{2} \frac{d^2}{dx^2} + V(x) + \sigma A^2 + \frac{C_n^2}{2} A^{2n} \right\} A = \omega A.
\end{equation}

Once the amplitude $A(x)$ and eigenvalue $\omega$ are determined (analytically or numerically, e.g., via self-consistent methods \cite{salerno-05}), the imaginary potential follows self-consistently from Eq.~(\ref{cmpxpot}).

Different choices of $n$ correspond to different effective nonlinearities. For instance, $n=1$ yields a cubic correction, while higher values generate higher-order nonlinear terms. These can be formally absorbed into a redefinition of the real potential:
\begin{equation}
V = \tilde{V} - \frac{C_n^2}{2} A^{2n},
\label{lin-n2}
\end{equation}
thus preserving the mapping structure. The approach can be extended to analytic functionals:
\begin{equation}
F(x) = \frac{1}{2} \sum_{n=0}^k C_n A^{n+2},
\label{complexg}
\end{equation}
as well as to spatial derivative expansions of the amplitude of the form:
\begin{equation}
F(x) = \sum_{n,m} C_{n,m} \frac{d^m A^{n+2}}{dx^m}.
\label{complexgen}
\end{equation}
The latter are particularly useful for modeling specific dynamical effects, such as the quadratic damping encountered
in harmonic traps. 

In all cases, the mapping guarantees that the resulting stationary solutions possess real eigenvalues (chemical 
potentials), despite the presence of complex potentials, ensuring their physical relevance in non-Hermitian photonics 
and dissipative quantum systems. 

A related reverse-engineering approach was considered in Ref.~\cite{Abdullaev2010}, where exact periodic waves, 
solitons, and breathers of the NLS equation with complex potentials were constructed analytically.

In the following sections, we extend the above mapping to time-dependent problems by applying it first to the linear
damped quantum harmonic oscillator and then to nonlinear dissipative dynamics in non-${\cal PT}$-symmetric complex 
potentials.


\section{The damped linear harmonic oscillator}
In this section, we consider the complex linear Schr\"odinger equation obtained from Eq.~(\ref{sys}) in the limit 
$\sigma=0$. This case provides a fundamental benchmark for understanding dissipative dynamics in non-Hermitian systems.

We first restrict the mapping to the simplest choice, $n=0$ in Eq.~(\ref{complex}), i.e., we take $F(x)=\frac{1}{2}C_0 
A^2(x)$. The resulting real eigenvalue problem (\ref{eigenval-prob}) reduces to a standard Schr\"odinger equation with 
a shifted eigenvalue $\omega=\Omega + C_0^2/2$. The corresponding imaginary potential and phase follow from the 
mapping as
\begin{equation}
W(x)=\frac{1}{A^2}\frac{dF}{dx}=C_0 \frac{A_x}{A}, 
\qquad 
\theta(x)=C_0 x.
\end{equation}

Assuming that the NESS  has the same functional form as the ground state of the harmonic oscillator, we take
\begin{equation}
A(x) = \left( \frac{\omega_0}{\pi} \right)^{1/4} e^{-\frac{\omega_0 x^2}{2}},
\label{gs_ha}
\end{equation}
which yields the imaginary potential
\begin{equation}
W(x) = C_0 \frac{A_x}{A} = -W_{0} x, 
\qquad 
W_{0}=C_0 \omega_0.
\label{w_undamp}
\end{equation}

This configuration corresponds to a ${\cal PT}$-symmetric system, where gain ($x<0$) and loss ($x>0$) are spatially balanced. The associated stationary solution is
\begin{equation}
\psi(x,t) = A(x)\, e^{i W_{0} x/\omega_0} e^{-i \omega t}.
\label{pisxt}
\end{equation}

This complex potential exactly supports the stationary state. However, despite its non-Hermitian character, it does not induce dissipative dynamics: a displaced initial wavefunction undergoes undamped oscillations. This reflects the unbroken ${\cal PT}$-symmetric phase, where the spectrum is real and no net dissipation occurs.

To describe genuinely damped dynamics, the mapping must be extended beyond the simple functional $F\propto A^2$. We therefore consider a functional that includes spatial derivatives (corresponding to $n=0$, $m=1$ in Eq.~(\ref{complexgen})):
\begin{equation}
F(x) = a_1 A^2(x) + \frac{a_2}{2} \frac{d}{dx} A^2(x).
\label{complexgg}
\end{equation}

We choose the real potential as a shifted harmonic oscillator,
\begin{equation}
V(x) = \frac{1}{2} \Omega_0^2 x^2 + 4 a_1 a_2 \omega_0 x,
\label{v_damped}
\end{equation}
with $\Omega_0^2 = \omega_0^2 (1 - 4 a_2^2)$. Under these conditions, the Gaussian profile (\ref{gs_ha}) solves the real eigenvalue problem with eigenvalue
\begin{equation}
\omega = \frac{1}{2}(\omega_0 + 4 a_1^2).
\end{equation}

The corresponding imaginary potential and phase are
\begin{eqnarray}
\label{w_damped}
W(x) &=& \frac{1}{A^2}\frac{dF}{dx}
= 2 a_2 \omega_0^2 x^2 - 2 a_1 \omega_0 x - a_2 \omega_0, \\
\theta(x) &=& 2 a_1 x - a_2 x^2.
\end{eqnarray}

\begin{figure}
\vskip -2.3 cm
\centerline{
\includegraphics[width=11.5cm,clip]{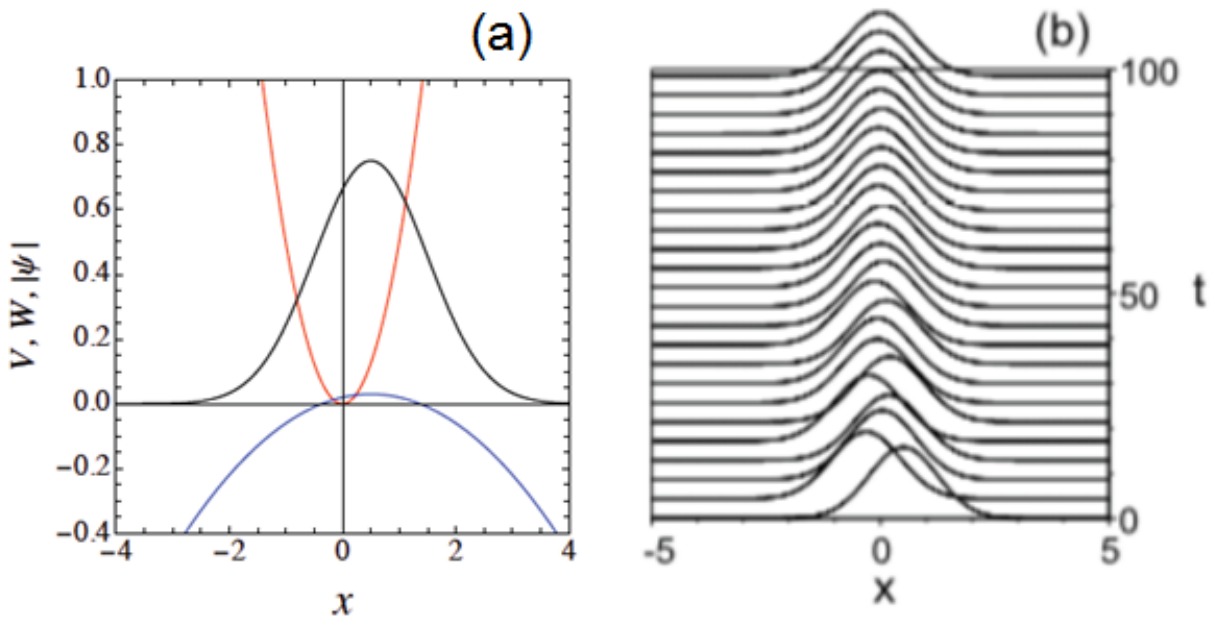}
}
\vspace{-2.5 cm}
\centerline{
\includegraphics[width=4.8 cm,clip]{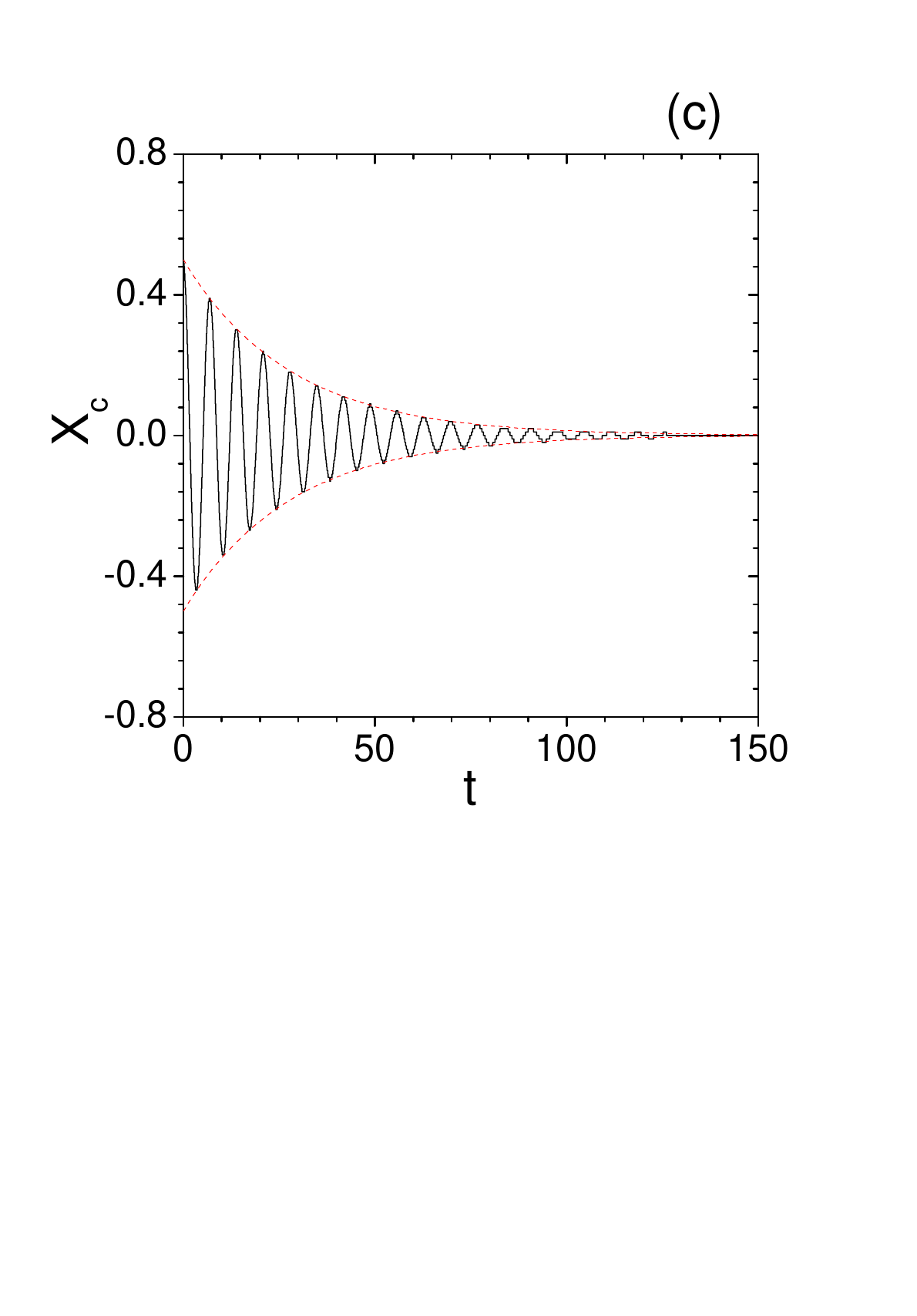}
\hspace{-1.2 cm}
\includegraphics[width=4.8 cm,clip]{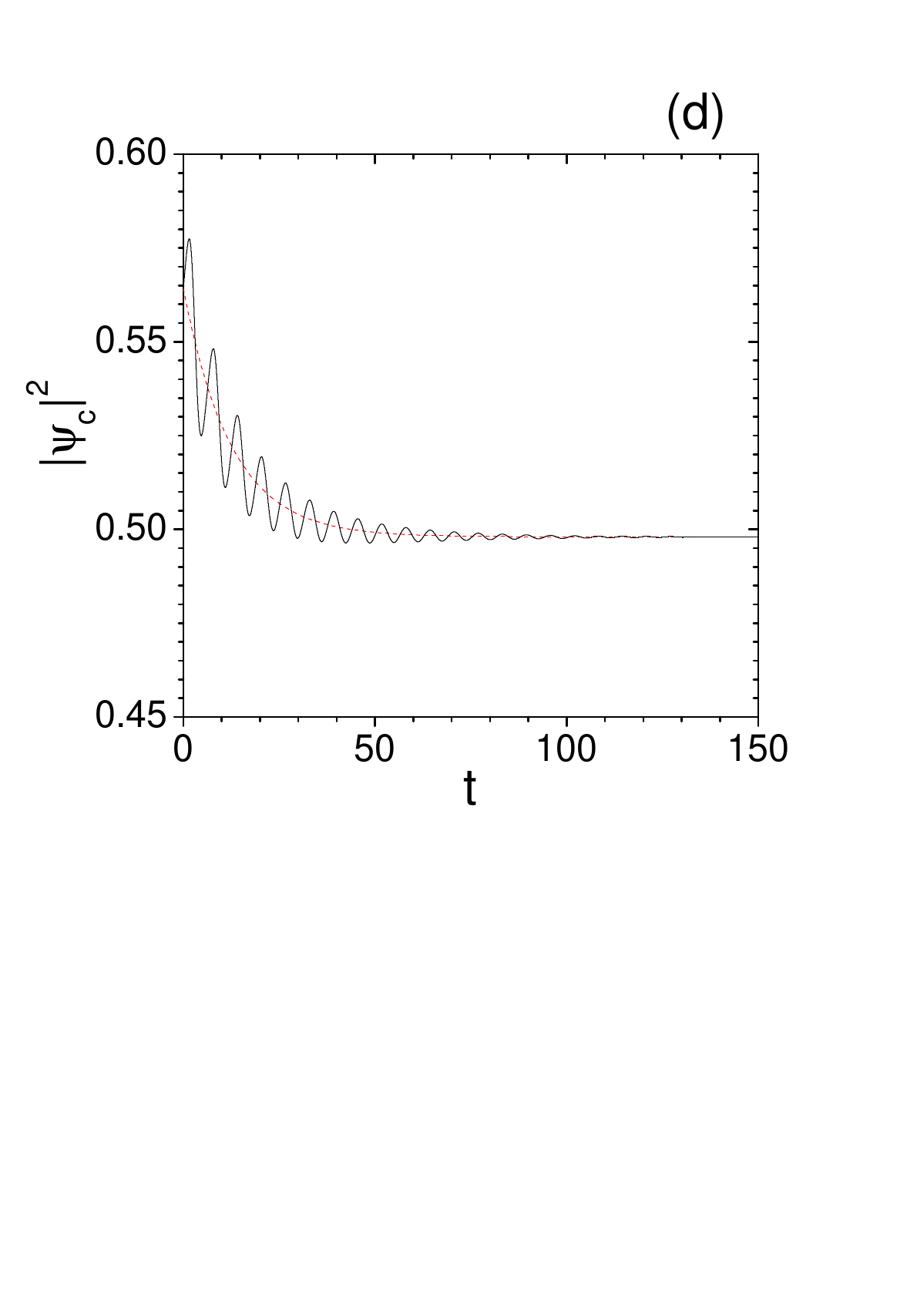}
}
\vskip -2.5 cm
\caption{
(a) Real (red line)  and imaginary (blue line) parts of the potential in Eq. (\ref{v_damped})
and (\ref{w_damped}) and ground state  wavefunction of the harmonic oscillator displaced by $x_0=0.5$. Other parameters are $\omega_0=1.0$, $a_{1} = a_2 = -0.02$.
(b) Time evolution of the density $|\psi(x,t)|^2$ for the initial state in (a).
 (c) Damped oscillations of the center of mass versus time. The dashed red  curves indicate the fit $x_c(t) \approx \pm x_0 e^{-\xi_c t}$ with  $\xi_c \approx 0.036$. (d) Peak density amplitude versus time relaxing toward the NESS (the red curve corresponds to Eq.~(\ref{fit})).
}
\label{fig1}
\end{figure}

In contrast to the ${\cal PT}$-symmetric case, the imaginary potential now includes a quadratic term ($\sim x^2$), which explicitly breaks ${\cal PT}$ symmetry and introduces a genuine dissipative channel. As shown in Figs.~\ref{fig1}(c,d), a displaced initial state exhibits damped oscillations, with the center of mass approximately following
\begin{equation}
x_c(t) \sim x_0 e^{-\gamma t}.
\end{equation}
At the same time, the peak density relaxes toward a stationary value, well described on average by

\begin{equation}
|\psi_c(t)|^2 \approx A_0 + \frac{e^{-\beta t} - 1}{\gamma},
\label{fit}
\end{equation}
(red curve in Fig.~\ref{fig1}(d)), with $\beta=0.08$, $\gamma=15.1$, and $A_0$ the density amplitude of the undamped ground state ($A_0 \approx 0.5642$).

It is remarkable that, in the linear case, the mapping provides a systematic framework for engineering complex potentials that accurately reproduce the dynamics of a damped quantum oscillator. In this sense, the mapping effectively extends beyond the stationary problem and captures the full dissipative evolution toward the NESS. This raises the question of whether the same construction remains valid in the presence of nonlinearity, which is addressed in the next section.


%
\begin{figure}
\centerline{
\includegraphics[width=4.2cm,clip]{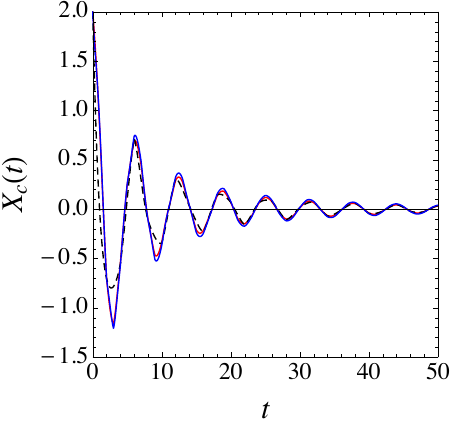}
\includegraphics[width=4.cm,clip]{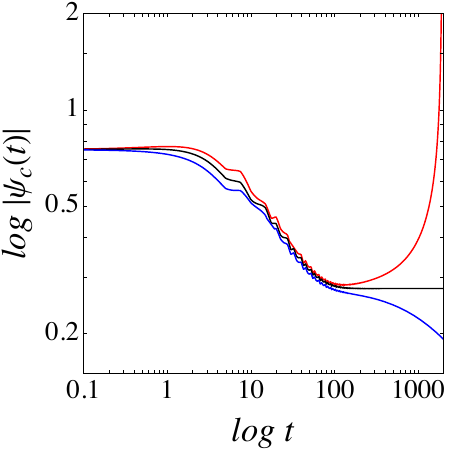}
}
\caption{
Time evolution of the center of mass (left panel) and peak amplitude (right panel) for the same complex potential used in the linear case, illustrating the effect of nonlinearity on the dissipative dynamics. The blue and red solid lines correspond to attractive ($\sigma=-1$) and repulsive ($\sigma=1$) interactions, respectively, while the dashed black line ($\sigma=0$) is shown for comparison. 
The logarithmic scale highlights the  slow evolution of the metastable states. Parameters are the same as in Fig.~\ref{fig1}, except for $x_0=2.0$, $a_1=-0.1$, and $a_2=-0.1$.
}
\label{fig2}
\end{figure}  

\section{Nonlinear dissipative dynamics}

\begin{figure}[t]
\vspace{-2.2cm}
\centerline{
\includegraphics[width=.69\columnwidth,clip]{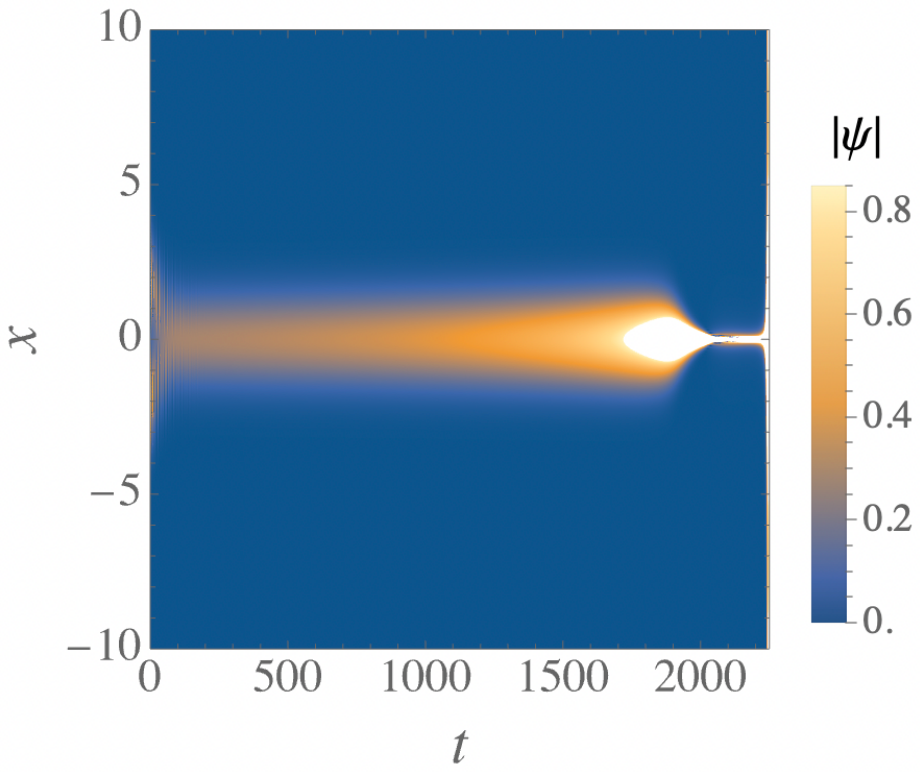}
\hspace{-1.6cm}
\includegraphics[width=.69\columnwidth,clip]{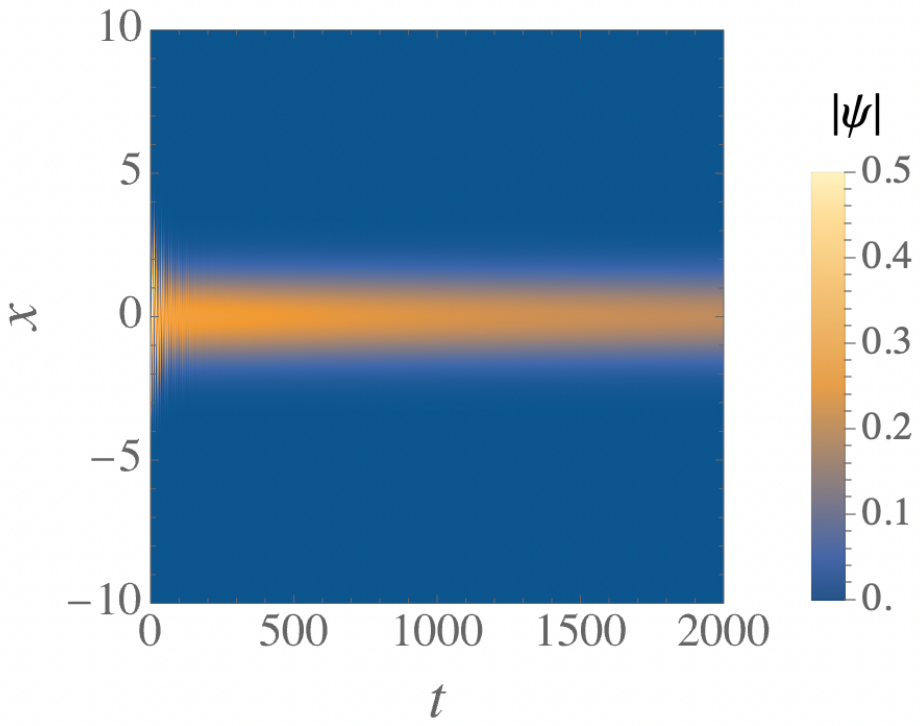}
}
\vskip -2.5cm
\caption{
Spatio-temporal evolution of the wave amplitude $|\psi(x,t)|$ in the nonlinear regime for $\sigma=-1$ (left panel) and $\sigma=1$ (right panel). The figures show the complete dissipative relaxation toward the trap center during the early stage of the dynamics, followed by the formation of long-lived metastable states. On longer timescales, the attractive case ($\sigma=-1$) undergoes collapse, whereas the repulsive case ($\sigma=1$) exhibits a slow decay of the wave amplitude. The complex potential and parameter values are the same as in Fig.~\ref{fig2}.
}
\label{fig3}
\end{figure}

We now address the nonlinear regime ($\sigma \neq 0$), using the same complex potential and initial conditions adopted in the linear case. This choice allows for a direct comparison and provides a stringent test of the mapping when applied to time-dependent nonlinear dynamics. The results are summarized in Figs.~\ref{fig2} and \ref{fig3}.

Figure~\ref{fig2} compares the time evolution of the center of mass and the peak amplitude for $\sigma=0,\pm1$. The left panel shows that, during the transient stage, the nonlinear dynamics closely follows the linear one: the wave packet undergoes damped oscillations toward the trap center with essentially the same decay rate. This indicates that the dissipative mechanism is primarily governed by the engineered complex potential, while nonlinear effects remain comparatively weak during the approach to the trap center.

The right panel of Fig.~\ref{fig2} reveals that at $t\approx100$, immediately after the dissipative transient, the nonlinear dynamics starts to deviate from the linear behavior. While the linear case reaches a constant peak amplitude corresponding to the formation of a NESS, the repulsive case ($\sigma=1$) displays a slow decrease of the amplitude, indicating a weak decay process. Conversely, the attractive case ($\sigma=-1$) exhibits a gradual growth of the peak amplitude (note the logarithmic scale in Fig.~\ref{fig2}), eventually leading to collapse on long timescales, as illustrated in the left panel of Fig.~\ref{fig3}.

The formation of metastable states at the trap center is also evident from the spatio-temporal evolution shown in Fig.~\ref{fig3}. In both attractive and repulsive cases, the system first relaxes toward a quasi-stationary localized state that remains remarkably close to the linear NESS for extended times before the residual instability eventually manifests itself.

This behavior highlights an important limitation of the mapping when applied to nonlinear dynamical problems. The mapping between real and complex Schr\"odinger equations is exact for stationary states. In time-dependent situations, however, a static potential design can accurately reproduce the dynamics only if the functional form of the wave remains unchanged, or approximately self-similar, during the dissipative evolution. This is precisely what occurs in the linear case. In the nonlinear regime, instead, the wave profile evolves dynamically under the combined action of dispersion, nonlinearity, gain, and loss. As a consequence, the exact gain--loss balance required for perfect NESS formation is no longer guaranteed and a small residual imbalance survives at the trap center.

It is remarkable, nevertheless, that the static complex potential design alone is capable of generating metastable states whose lifetimes largely exceed the characteristic dissipative timescale of the transient dynamics in both attractive and repulsive regimes. In the next section, we show that these residual long-term instabilities can be efficiently suppressed by means of nonlinear management, leading to the stabilization of the NESS.
\begin{figure}[t]
\centerline{
\includegraphics[width=.5\columnwidth,clip]{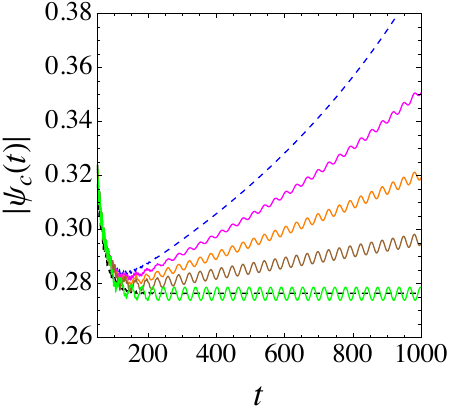}
\hspace{-.25cm}
\includegraphics[width=.5\columnwidth,clip]{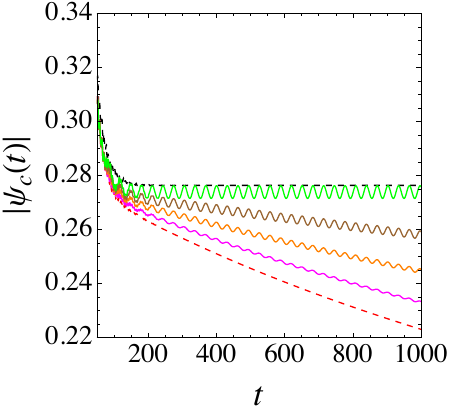}
}
\caption{
Time evolution of the peak amplitude for attractive, $\sigma_0=-1$ (left  panel), and repulsive,  $\sigma_0=1$ (right panel) in the presence of NM nonlinear (continuous curves) with different values of $\gamma$. The continuus curves, with magenta, orange, brown, green, colors correspond in both panels to the amplitudes $\gamma= 0.5, 1.0, 1.5, 2.0.$, respectively.  The dashed blue, black, and red lines  refer to the value $\gamma=0$ (i.e. absence of NM) and coincide with the curves of the same color depicted  in Fig.~\ref{fig3} here plotted is linear scale. The modulation frequency is $\Omega=0.2$ and the activation times are $t_0 = 100$ and  $t_0=80$ for left and right panels, respectively. All other parameters are the same as in Fig.~\ref{fig2}.
}
\label{fig4}
\end{figure}
%

\section{Nonlinear management and NESS stabilization}

The fact that the residual imbalance, unavoidable within a static potential design, depends on the sign of the nonlinearity naturally suggests a stabilization mechanism for the NESS based on nonlinear management (NM), i.e., time-periodic modulations of the interaction term \cite{malomed2006soliton,centurion2006nonlinmanag}.

NM has been extensively investigated as a tool for controlling collapse in attractive multidimensional nonlinear systems \cite{abdullaev2003nonlinearity,torres2000dispersion}. However, to our knowledge, its application to the stabilization of non-equilibrium stationary states in nonlinear open systems has received much less attention \cite{diehl2008quantum,disscooling2022}. We stress that the collapse and decay processes displayed in Fig.~\ref{fig3} are not related to dimensionality effects, but rather originate from the residual mismatch between gain, loss, and nonlinear interactions that remains in the static potential design. A temporal modulation of the nonlinearity can therefore provide an additional dynamical mechanism capable of compensating for this imbalance and stabilizing the NESS.

To test this idea, we introduce the NM protocol
\begin{equation}
\sigma(t) =
\begin{cases}
\sigma_0, & t \le t_0, \\
\sigma_0\left( 1 - \gamma + \frac{\gamma}{2} \left[1 + \cos(\Omega (t-t_0))\right]\right), & t > t_0,
\end{cases}
\label{NM}
\end{equation}
where $\sigma_0=\pm1$, $t_0$ denotes the activation time, $\Omega$ is the modulation frequency, and $\gamma$ controls the modulation amplitude. The management is activated after the damped oscillatory stage, when the wave packet has already approached the trap center and the metastable regime is formed.

The results are summarized in Fig.~\ref{fig4}, where we report the time evolution of the peak amplitude for different values of $\gamma$ in both the attractive and repulsive cases. As $\gamma$ increases within the interval $0\le\gamma\le2$, the amplitude growth observed in the attractive case is progressively suppressed until, at $\gamma=2$, the state becomes effectively stationary. A similar stabilization against decay occurs in the repulsive case.

Remarkably, in both cases the stabilized NESS is obtained when the nonlinearity oscillates between $\pm1$, corresponding to a vanishing time-averaged interaction. In this regime, the average peak amplitude becomes very close to that of the linear NESS discussed in Sec.~III. This indicates that the modulation dynamically compensates for the residual nonlinear imbalance responsible for the slow collapse or decay observed in the unmanaged dynamics.

The modulation frequency used in Fig.~\ref{fig4} is intentionally chosen sufficiently small to make the residual oscillations of the peak amplitude visible on the expanded scale of the figure. The oscillation period coincides with the modulation period $2\pi/\Omega$, while their amplitudes remain very small even for the largest values of $\gamma$. 

We have verified that increasing the modulation frequency $\Omega$, while keeping all other parameters fixed, does not qualitatively modify the dynamics reported in Fig.~\ref{fig4}. In particular, the average peak-amplitude evolution remains essentially unchanged, whereas the small oscillations become progressively denser due to the reduced modulation period. This demonstrates that the nonlinear response of the system to the NM remains active even in the regime of vanishing time-averaged nonlinearity.

We also note that stabilization does not require the limiting case $\gamma=2$. A broad interval $0<\gamma<2$ already produces substantial  increase of the metastable lifetime, effectively pushing the onset of collapse or decay beyond experimentally relevant time scales.
 
\begin{figure}[t]
\vspace{-2.2cm}
\centerline{
\includegraphics[width=.73\columnwidth,clip]{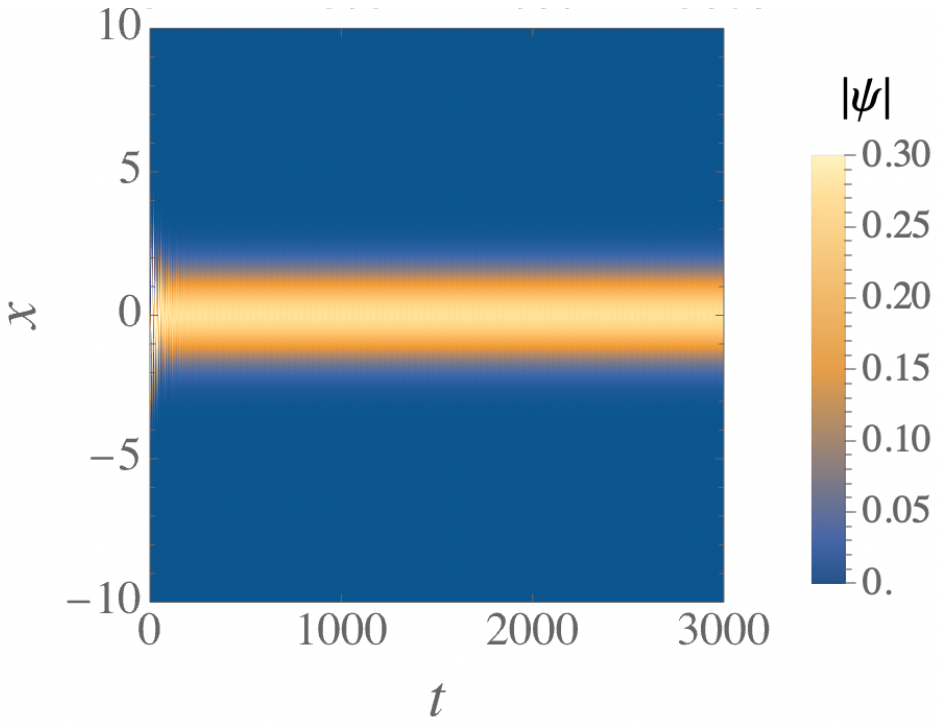}
\hspace{-2.5cm}
\includegraphics[width=.73\columnwidth,clip]{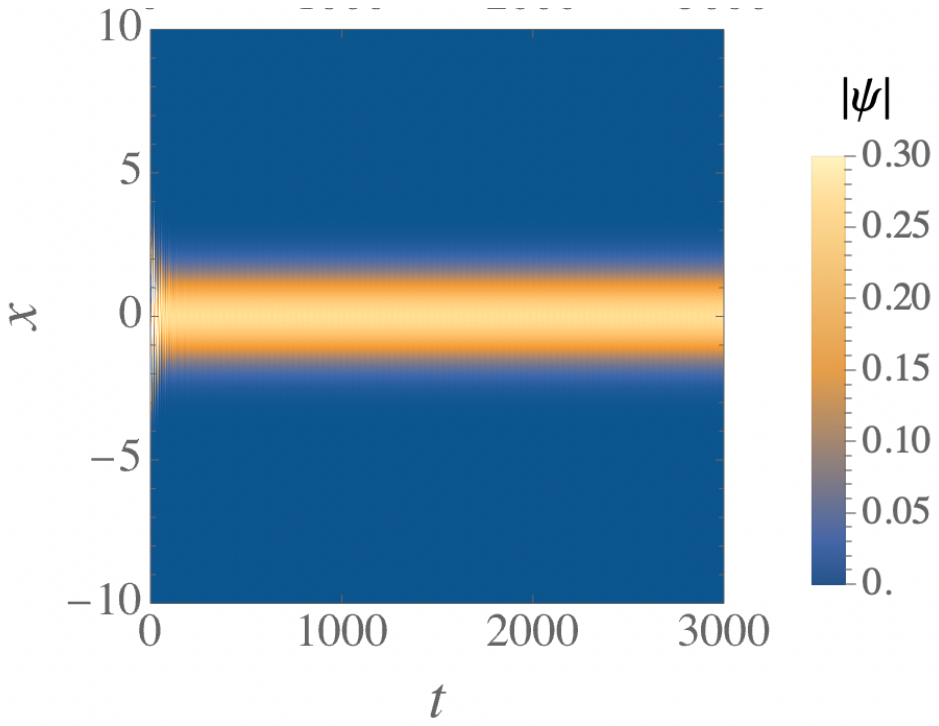}
}
\vskip -2.5cm
\caption{
Spatio-temporal evolution of the wave  amplitude $|\psi(x,t)|$ in the presence of the NM protocol in Eq. (\ref{NM})
with $\gamma=-2$ and $\sigma_0=-1,\, t_0=100$ (left panel), $\sigma=1,\,  t_0=75 $ (right panel). All other parameters are fixed as in Fig.~\ref{fig3}. 
}
\label{fig5}
\end{figure}

The effectiveness of the NM protocol is further illustrated in Fig.~\ref{fig5}, where the spatio-temporal evolution of the wave amplitude is reported for the same cases shown in Fig.~\ref{fig4}, but with the NM protocol of amplitude $\gamma=2$ activated. In both attractive and repulsive cases, the metastable configuration evolves into a robust NESS characterized by a nearly stationary localized profile and a dynamically sustained balance between gain, loss, and nonlinear interactions.

%

\section{Conclusions and Experimental Perspectives}

In this work, we have investigated the dissipative dynamics of linear and nonlinear waves in harmonic traps using engineered complex non-Hermitian potentials. By exploiting a mapping between real and complex Schr\"odinger equations, we have shown that it is possible to design complex potentials that induce controlled damping toward the trap center.

In the linear case, this mapping provides an exact description not only of stationary states but also of the full time-dependent dynamics. Consequently, an initially displaced wave packet undergoes damped oscillations and relaxes toward a NESS localized at the trap center. This demonstrates that complex potential engineering offers a systematic framework for realizing dissipative quantum oscillators within a non-Hermitian Schr\"odinger setting.

In the nonlinear regime, the same engineered potential governs the transient dynamics, which closely follows the linear case. However, the asymptotic behavior is fundamentally different: instead of converging to a true stationary state, the system evolves toward a long-lived metastable configuration. This reveals that, while the mapping remains exact for stationary solutions, the inherent dynamical feedback of the nonlinearity prevents a perfect power balance under a static potential design. As a result, a residual gain--loss imbalance persists, leading to slow decay in the repulsive case or gradual focusing and eventual collapse in the attractive case. We have shown that stability can be restored by combining mapping-derived potentials with active control of the nonlinearity. Specifically, a time-dependent modulation of the interaction strength suppresses internal excitations and leads to the formation of robust NESS, highlighting the importance of dynamical control strategies in non-Hermitian nonlinear systems.

From an experimental perspective, the proposed scenario is directly accessible in both photonic and atomic systems. In nonlinear optics, the real and imaginary parts of the potential can be engineered through refractive index modulation and spatially distributed gain/loss, while Kerr nonlinearities provide access to the interacting regime. In Bose--Einstein condensates, harmonic confinement is naturally available, and the required complex landscapes can be implemented using localized atom removal (via electron beams or optical dissipators) and feeding techniques, whereas the interaction strength remains finely tunable via Feshbach resonances.

A key advantage of this framework is that the same engineered complex potential can be used to explore both linear and nonlinear regimes without modification. This enables a direct comparison between exact dissipative stabilization and metastable nonlinear dynamics. In particular, the long-lived localized states observed in the nonlinear regime, together with their slow decay or focusing signatures and their full stabilization via NM, represent a robust experimental hallmark of the interplay between nonlinearity and non-Hermiticity.

\bibliography{A-referenze}

\end{document}